\documentclass[conference]{IEEEtran}
\IEEEoverridecommandlockouts
\usepackage{cite}
\usepackage{amsmath,amssymb,amsfonts}
\usepackage{algorithmic}
\usepackage{graphicx}
\usepackage{textcomp}
\usepackage{xcolor}
\usepackage{comment}
\usepackage{url}
\usepackage{booktabs}
\usepackage{float} 
\usepackage{algorithm}
\usepackage{algorithmic}
\usepackage[caption=false, font=footnotesize]{subfig}
\def\BibTeX{{\rm B\kern-.05em{\sc i\kern-.025em b}\kern-.08em
    T\kern-.1667em\lower.7ex\hbox{E}\kern-.125emX}}

\begin{document}
\title{Bridging Biological Hearing and Neuromorphic Computing: End-to-End Time-Domain Audio Signal Processing with Reservoir Computing\\}
%Time domain Audio Signal Processing Using Reservoir Computing\\}
%\author{***REDACTED for double blind review***}

%\begin{comment}

\author{\IEEEauthorblockN{1\textsuperscript{st} Rinku Sebastian}
\IEEEauthorblockA{ \textit{University of York.}\\
York, United Kingdom. \\
rinku.sebastian@york.ac.uk}
\and
\IEEEauthorblockN{2\textsuperscript{nd} Simon O'Keefe}
\IEEEauthorblockA{\textit{University of York.}\\
York, United Kingdom. \\
simon.okeefe@york.ac.uk}
\and
\IEEEauthorblockN{3\textsuperscript{rd} Martin A. Trefzer}
\IEEEauthorblockA{\textit{University of York.}\\
York, United Kingdom. \\
martin.trefzer@york.ac.uk}
}
%\end{comment}
\maketitle

\begin{abstract}
Despite the advancements in cutting-edge technologies, audio signal processing continues to pose challenges and lacks the precision of a human speech processing system. To address these challenges, we propose a novel approach to simplify audio signal processing by leveraging time-domain techniques and reservoir computing. Through our research, we have developed a real-time audio signal processing system by simplifying audio signal processing through the utilization of reservoir computers, which are significantly easier to train. 

Feature extraction is a fundamental step in speech signal processing, with Mel Frequency Cepstral Coefficients (MFCCs) being a dominant choice due to their perceptual relevance to human hearing. However, conventional MFCC extraction relies on computationally intensive time-frequency transformations, limiting efficiency in real-time applications. To address this, we propose a novel approach that leverages reservoir computing to streamline MFCC extraction. By replacing traditional frequency-domain conversions with convolution operations, we eliminate the need for complex transformations while maintaining feature discriminability. We present an end-to-end audio processing framework that integrates this method, demonstrating its potential for efficient and real-time speech analysis. Our results contribute to the advancement of energy-efficient audio processing technologies, enabling seamless deployment in embedded systems and voice-driven applications. This work bridges the gap between biologically inspired feature extraction and modern neuromorphic computing, offering a scalable solution for next-generation speech recognition systems.
\end{abstract}
keywords: {Reservoir computing, Audio signal processing, MFCC. Time domain audio feature extraction}

\section{Introduction}
 Effective audio processing is increasingly essential for many modern technologies, including communications, computerized speech transcription and translation, speaker verification, hearing aids, etc~\cite{Anusuya_2009-uj}. Audio signals are temporal signals. In other words, the characteristics of a signal change significantly over time. This temporal complexity makes audio signal processing particularly challenging.

  The majority of contemporary audio processing entails translating audio signals to frequency-domain. Most of these translations eliminate the time information in the signal and introduce limitations like the irreversible loss of precise time-localized features during Fourier transformations, which significantly hinders tasks that require precise temporal alignment, and significant computational overhead from repeated domain conversions. These drawbacks limit neural networks' capacity to extract the best representations straight from unprocessed waveforms, and they also require significant resources for pre-processing instead of core model optimization.  
 
A commonly used approach is to extract Mel Frequency Cepstral Coefficients (MFCC) from an audio stream. The standard MFCC extraction process involves multiple computationally demanding stages: pre-emphasis and framing of the input signal, followed by Fourier transformation through FFT, application of mel-scale filter-banks, logarithmic compression, and finally discrete cosine transform to produce the cepstral coefficients. Research indicates this conventional approach requires significantly more computational resources~\cite{He2022-cy}. More than half of the processing time is spent on the FFT and mel-filterbank procedures alone. The repeated domain conversions not only increase latency but also create memory bottlenecks, particularly for real-time applications. This complexity has motivated us to explore time-domain alternatives that could potentially replicate MFCC-like features, so that the process is simplified and the burden of calculating extra time dependent coefficients is avoided.

A critical area of exploration is the implementation of RC in physical hardware, which could revolutionize energy-efficient audio processing. By simplifying audio signal processing for these hardware-compatible reservoirs, researchers can contribute to the development of next-generation, low-power devices capable of human-like auditory performance. The ongoing investigation into RC’s potential in audio signal processing aims to address the limitations of current technologies while paving the way for innovative solutions. By combining the efficiency of reservoir computing with advancements in neuromorphic engineering, future systems could achieve real-time, energy-efficient audio processing that rivals biological systems.

Our work investigates reservoir computing (RC) as an energy-efficient alternative for audio processing tasks. Unlike conventional methods, reservoir's dynamical systems approach can directly process time-domain waveforms while maintaining the temporal precision essential for speaker identification and the spectral sensitivity needed for digit recognition. This eliminates the need for costly frequency-domain conversions and handcrafted feature extraction.
 
\section{Reservoir computing}
Reservoir computing is a machine learning approach inspired by biological systems. It operates within a computational framework derived from recurrent neural network (RNN) principles, where input signals are transformed into higher-dimensional representations through the dynamics of a fixed, non-linear system called a reservoir. In this paradigm, the reservoir acts as a black box—its internal connections are randomly initialized and remain untrained. Only a simple readout mechanism is trained to interpret the reservoir's state and produce the desired output~\cite{Stepney2024-dx}. Unlike traditional RNNs, where all weights are adjusted during training, reservoir computing typically employs efficient regression techniques solely to optimize the readout layer while keeping the reservoir's dynamics unchanged.

\begin{figure}[h]
    \centering
    \includegraphics[width=\columnwidth]{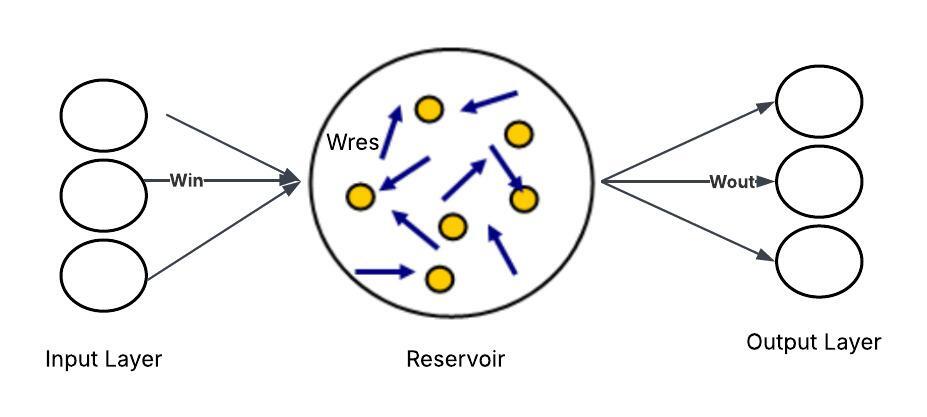}
    \caption{Topology of Reservoir computer}
    \label{fig:enter-labelr1}
\end{figure}

Since RNN development is sluggish and challenging, in 2001 Wolfgang Maass and Herbert Jaeger independently suggested Liquid State Machines~\cite{Maass2002-tg} and Echo State Networks~\cite{Jaeger_2010-pc} as fundamentally new approaches to RNN design and training. Reservoir Computing is a term that has since been coined to refer to these methods. It has roots in computational neuroscience~\cite{Dominey2004ComplexSS} and later consequences in machine learning as the Backpropagation-Decorrelation~\cite{Steil2004-kk} learning rule (RC).  Figure~\ref{fig:enter-labelr1} shows a classical reservoir computer. The RC framework consists of three core components. An input layer that is randomly connected to each of the N reservoir nodes receives the input. The reservoir itself is left untrained since the connections and weights between its nodes are fixed and selected at random.  An output layer reads out the transient dynamical response of the reservoir using linear weighted summing of the node states \cite{Wringe2025-xe}.

The input signal $ \mathbf{u}(t) \in \mathbb{R}^M $is mapped into the reservoir space via:
\begin{equation}
\mathbf{x}_{in}(t) = \mathbf{W}_{in} \mathbf{u}(t)
\end{equation}

where $\mathbf{W}_{in} \in \mathbb{R}^{N \times M}$ projects inputs $\mathbf{u}(t) \in \mathbb{R}^M$ into the $N$-dimensional reservoir space.

The reservoir state $\mathbf{x}(t) \in \mathbb{R}^N$ evolves as:

\begin{equation}
\mathbf{x}(t) = f\left(\mathbf{W}_{res} \mathbf{x}(t-1) + \mathbf{x}_{in}(t) + \mathbf{b}\right)
\end{equation}

with:
\begin{itemize}
\item $\mathbf{W}_{res} \in \mathbb{R}^{N \times N}$: Sparse recurrent weight matrix (spectral radius $\rho < 1$)
\item $f(\cdot)$: Nonlinearity (typically $\tanh$)
\item $\mathbf{b} \in \mathbb{R}^N$: Optional bias
\end{itemize}

The output is computed by:

\begin{equation}
\mathbf{y}(t) = \mathbf{W}_{out} \mathbf{x}(t), \quad \mathbf{W}_{out} \in \mathbb{R}^{P \times N}
\end{equation}

where $\mathbf{W}_{out}$ is the only trained component, typically learned via ridge regression:

The drawbacks of gradient-descent RNN training are avoided by the RC paradigm. This made it much easier to use RNNs in real-world applications and outperformed traditional fully trained RNNs in many tasks~\cite{Lukosevicius2009-ib}. Reservoir systems process inputs as a continuous stream, where the reservoir’s high-dimensional dynamics naturally mix past and present information. Each new input perturbs the reservoir’s state, which retains echoes of previous inputs due to recurrent connections.  The readout layer then extracts relevant features from this rich, evolving state.

The utility of neural networks in the audio signal processing domain has been explored for a long time since it is a complex task. In the reservoir framework, since the training is limited to the readout part, the burden of training is reduced. Also, Interference between the tasks is also minimized if we are performing multiple task by training multiple readouts on the same reservoir. It is possible to solve several tasks with a single input by adding multiple readouts to a single reservoir. So multitasking can be efficiently or effectively employed using reservoirs. The echo state property of a reservoir gives the system memory so that it can process time series. This fading memory property allows RC to model temporal dependencies without explicit back-propagation through time and prevents the system from saturation. Furthermore, the reservoir has the  ability to perform non-linear transformations. All these qualities of a reservoir show that it is a suitable fit for temporal signal processing~\cite{Grezes2025-oo}.

\section{Audio signal processing}

Audio signal analysis involves characterizing, modeling, and interpreting sound data by uncovering underlying patterns and relationships between signals. This process applies to diverse acoustic inputs, including speech, music, and environmental sounds. Modern advancements in signal processing and machine learning (ML) have significantly enhanced audio classification and pattern recognition. Any ML algorithm's performance is based on the features used for training and testing. Thus, feature extraction is one of the most important processes in a machine learning process~\cite{Sharma2020-pb}.

Feature extraction transforms raw audio waveforms into lower-dimensional, information-rich representations while preserving essential characteristics. Since directly analysing high-resolution audio data is computationally impractical, this step identifies and retains only the most relevant attributes for the task. Effective features capture a signal's distinctive properties in a condensed form, enabling efficient downstream processing. This is due to the fact that processing all of the information in the acoustic signal would be intractable, and some of it is not relevant for the purpose~\cite{Rashid18}. The choice of features critically impacts model performance, making this stage fundamental in audio-based machine learning pipelines.

The following section describes Mel Frequency Cepstral Coefficient in detail.

\subsection{MFCC}
Mel-Frequency Cepstral Coefficients (MFCCs) have become a fundamental feature extraction technique in speech processing due to their ability to closely match human auditory perception. By converting linear frequency scales to the non-linear Mel scale - which more accurately represents how humans hear sounds, especially at higher frequencies - MFCCs provide a perceptually relevant representation of audio signals. This psycho-acoustic transformation makes MFCCs particularly valuable for speech and speaker recognition systems, where modelling human-like perception improves performance.

The MFCC extraction process involves several key transformations of the audio signal's power spectrum. First, the frequency axis is warped to the Mel scale, creating a representation that emphasizes perceptually important frequency ranges. Then, a logarithmic compression and discrete cosine transform are applied to produce de-correlated coefficients that compactly represent the signal's spectral envelope. The resulting coefficients correspond to equally spaced frequency bands on the Mel scale, providing an efficient yet perceptually meaningful parameterization of the sound's spectral characteristics. This combination of mathematical processing and psycho-acoustic principles has made MFCCs one of the most successful and enduring feature sets in speech technology.

MFCCs are commonly derived as follows:
\begin{itemize}
    \item 	Step 1: Take the Fourier transform of (a windowed excerpt of) a signal.
\item	Step 2: Map the powers of the spectrum obtained above onto the Mel scale, using triangular overlapping windows or alternatively, cosine overlapping windows.
\item	Step 3: Take the logs of the powers at each of the Mel frequencies.
\item Step 4: Take discrete cosine transform of the list of Mel log powers.
\item	The MFCCs are the amplitudes of the resulting spectrum
\end{itemize}
\begin{figure}
    \centering
    \includegraphics[width=.40\columnwidth]{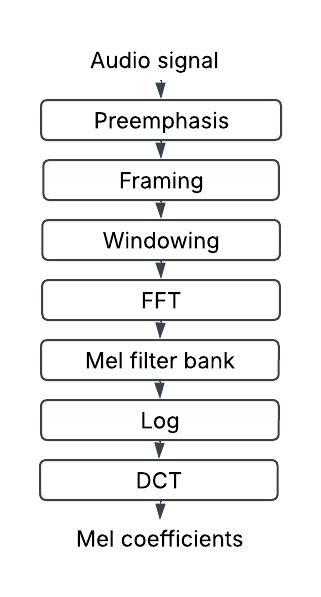}
    \caption{MFC extraction }
    \label{fig:my_labelmfc}
\end{figure}
\textbf{Framing and windowing:} 
Since MFCC analysis relies on spectral information, the audio signal must first be converted from the time domain to the frequency domain. Speech signals are typically considered stationary only in short segments, with their periodicity characteristics varying based on duration: segments shorter than 30 ms can be treated as periodic, while those between 30-200 ms exhibit uncertain periodicity, and longer segments are non-periodic. To capture these short-term stationary properties, the signal is divided into 20-30 ms frames with a 10 ms overlap between consecutive frames ~\cite{Niewiadomy_2008-zy}. This overlapping frame approach ensures that each speech sound appears centred within at least one analysis window. A 20 ms window duration provides an optimal balance - it's sufficiently long to capture key spectral features while maintaining good temporal resolution. Before applying the Discrete Fourier Transform (DFT), each frame is multiplied by a smoothing window function (typically Hamming or Hanning) to taper the signal at the frame edges. This windowing operation serves three important purposes: it enhances harmonic components, smooths frame boundaries, and minimizes edge artifacts that could distort the spectral analysis.

\textbf{DFT spectrum:} Each windowed frame is converted into frequency spectrum by applying DFT.
\begin{equation}
    X(k)=\sum_{n=0}^{N-1} x(n)*e^ {-j2\pi nk/N} 
\end{equation}

\textbf{Mel spectrum:}  Mel spectrum is computed by passing the Fourier transformed signal through a set of band-pass filters known as Mel-filter bank. A Mel is a unit of measurement of how loudness is perceived by the human ear. Since the human auditory system reportedly does not detect pitch linearly, it does not correspond linearly to the tonal frequency physically present in the sound. The frequency spacing for the Mel scale is roughly linear below 1~kHz and logarithmic above 1~kHz. Mel can be approximated by physical frequency using the formula
\begin{equation}
    f_{Mel}=2595 \log_{10}(1+f/700)
\end{equation}
Where f denotes the physical frequency in Hz, and $f_{Mel}$ denotes the perceived frequency
\begin{figure}
    \centering
    \includegraphics[width=0.5\textwidth]{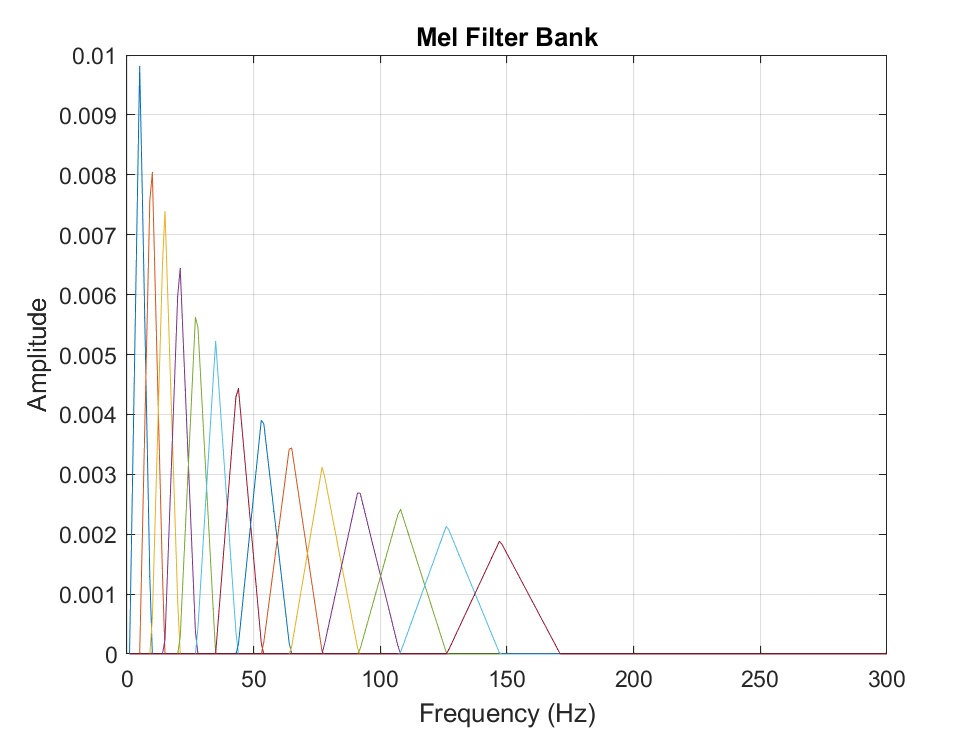}
    \caption{Mel filter bank}
    \label{fig:my_label21}
\end{figure}
%Both the frequency domain and the time domain are capable of representing filter banks.
Filter banks are typically built in the frequency domain for MFCC calculations. On the frequency axis, the centre frequencies of the filters are typically uniformly spaced. However, the warped axis, in accordance with the non-linear function provided in equation (5), is implemented in order to match the human ear's perception~\cite{Rao2017-hz}. The filter bank typically consists of overlapping triangular filters~\cite{Sebastian2026-eo}. Figure~\ref{fig:my_label21} shows the generated Mel filter bank for 1024 point FFT transform, where the number of filters is 25, minimum frequency is 0 Hz, maximum frequency is 4000 Hz and sampling frequency is 8 kHz. The algorithm generating MFCCs  creates the filter bank before processing is done, because filter bank parameters are constant. The frequency spectrum of the signal(i.e., X(k) from equation (4) is multiplied with the filter bank to obtain mel frequency spectrum. Thus mapping the power-spectrum of the signal on to the Mel scale.  

\textbf{Discrete cosine transform (DCT):}
The smooth nature of the vocal tract creates inherent correlations between adjacent frequency bands' energy levels. To address this and extract meaningful features, the Mel-frequency coefficients undergo two key processing steps. First, the Mel spectrum is converted to a logarithmic scale to better represent human loudness perception and normalize amplitude variations. Then, the Discrete Cosine Transform (DCT) is applied to de-correlate the spectral components, transforming them into cepstral coefficients. In the resulting cepstral domain, low quefrency regions capture the vocal tract's formant structure while higher quefrency components correspond to pitch information. Since the first few coefficients typically contain most of the spectral envelope information relevant for speech recognition, higher-order coefficients representing finer details can often be discarded. This selective coefficient retention makes the system more robust while maintaining speech intelligibility, as it focuses on the most perceptually significant features and reduces sensitivity to pitch variations and noise. The combination of logarithmic compression and DCT transformation effectively concentrates the speech signal's essential characteristics into a compact set of de-correlated parameters ideal for pattern recognition tasks

Finally, MFCC are calculated as
\begin{equation}
    c(n)=\sum_{m=0}^{M-1} \log_{10} (s(m))cos(\pi n(m-0.5)/M)
\end{equation}
n=0, 1,2....C-1.
where, M is total number of triangular Mel weighting filters and c(n) are the cepstral coefficients, and C is the number of MFCCs. MFCC systems use only 8–13 cepstral coefficients. The zeroth coefficient is often excluded since it represents the average log-energy of the input signal, which only carries small amount of speaker-specific information. \cite{Rao2017-hz}

In the final processing stage, the logarithmic Mel spectrum undergoes transformation back to a time-domain representation through the Discrete Cosine Transform (DCT), yielding the Mel Frequency Cepstral Coefficients (MFCCs). The DCT effectively de-correlates the spectral components, producing a compact cepstral representation that preserves the signal's essential spectral characteristics while discarding redundant information. The resulting MFCCs provide an optimal time-domain representation of the signal's local spectral properties for the analysed frame, capturing the most perceptually relevant features of the original speech signal in a form suitable for pattern recognition and machine learning applications. This cepstral transformation completes the MFCC feature extraction pipeline, converting the perceptually-warped frequency analysis into a robust parametric representation of the acoustic signal~\cite{Tiwari_2010-xe}.

\textbf{Deltas and Delta-Deltas:} Deltas and Delta-Deltas are also known as differential and acceleration coefficients. Only the power spectral envelope of a single frame is described by the MFCC feature vector, but speech also contain information about dynamics, i.e., the trajectory of the MFCC coefficients over time. Adding the MFCC trajectories to the original feature vector after computing them, significantly improves automatic speech recognition performance. The benefit of Delta features is that they are used to represent the temporal information. To calculate the delta coefficients, the following formula is used.
\begin{equation}
    d_t = \frac{\sum_{n=1}^N n (c_{t+n} - c_{t-n})}{2 \sum_{n=1}^N n^2}
\end{equation}
where $d_t$ is a delta coefficient from frame t computed in terms of the static coefficients $c_{t-n}$ to $c_{t+n}$. n is usually taken to be 2. By taking the derivative of Delta features, Delta-Delta features are extracted~\cite{Singh_2016-vk}.

\section{Motivation}
Figure \ref{fig:my_labelmfc} shows the MFCC extraction. The primary motivation behind these steps is mathematical simplification. Convolution in the time domain corresponds to multiplication in the frequency domain, which further reduces to addition in the log-frequency domain. But,they inadvertently introduce complexity into the MFCC computation.  Given that reservoir computing excels at time-domain processing, we explored simplifying MFCC extraction using a reservoir-based approach. 
In addition, in order to recognize speech better, we need to understand the dynamics of the power spectrum, i.e., the trajectories of MFCCs over time. For estimating these, we need delta and delta-delta coefficients to be calculated. This further complicates the speech signal processing.

\section{Methodology}
This work presents a comprehensive analysis of reservoir computing (RC) for real-time audio processing, with a focus on direct time-domain operation. We propose a novel dual-role RC architecture that simultaneously functions as: a feature extractor replacing conventional mel-frequency processing stages and a temporal pattern classifier. Our approach eliminates the need for explicit frequency-domain transformations, instead learning equivalent spectral representations directly from raw waveforms through the reservoir’s inherent dynamics. 

We have pre-processed the speech signals by extracting information using the Mel frequency Cepstral Coefficient (MFCC). We extracted the first 14 MFCC coefficients from the speech signal, which represent the short-term spectral features of the audio. These coefficients capture the shape of the vocal tract and are commonly used for speech signal processing. We have used the TI-46 dataset, which consists of eight female speakers uttering digits 0~to~9, 10 times each. Additionally, we  used the Audio-Mnist data set, which consists of 60 speakers uttering the numbers 0–9, 50 times each. To confirm the functionality of the system, we conducted experiments that involved both speaker and digit recognition.

%To begin with, We have separated the TI-46 data set into 5 partitions where each contains samples of all 8 speakers uttering all 10 digits twice for the purpose of conducting 5-fold cross validation.

\subsection{ Reservoir as a classifier}
This is to test the ability of a RC to classify audio based on features that have been extracted conventionally.

\begin{figure}
    \centering
      \subfloat[Reservoir as classifier]{\includegraphics[width=0.9\columnwidth]{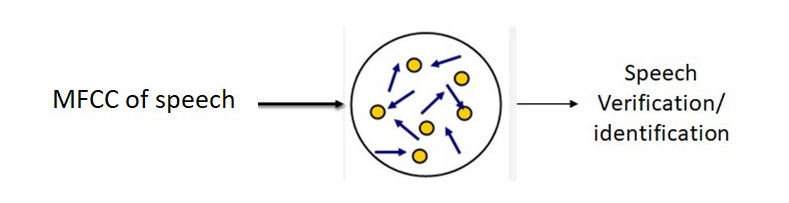}}\\
      \subfloat[Reservoir as feature extractor and classifier]{\includegraphics[width=\columnwidth]{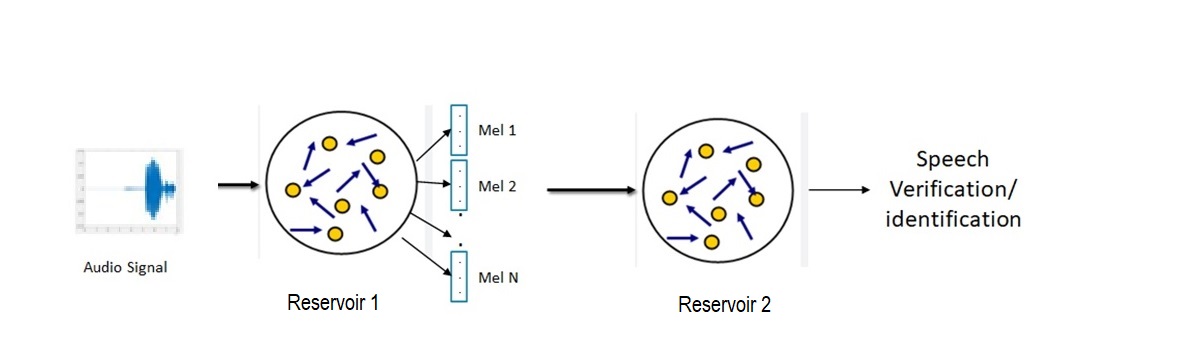}}
    \caption{A comparison between the conventional approach (a) where the reservoir is used as a classifier after complex pre-processing by different means and the approach proposed here (b) where RC is used as an end-to-end audio processing concept.}
    \label{fig:mylabel_RFC}
\end{figure}

%\subsubsection{Experiment Setup}
Figure~\ref{fig:mylabel_RFC}(a) shows the reservoir as a classifier. The reservoir we used for classification is made up of 400 nodes with a sparsity of 80\%, indicating that the connections were established with a probability of 20\%.The leakage rate is 0.3 and washout is 50. We separated each digit, calculated the MFCC of each digit using Matlab, and concatenated the MFCC of each digit, representing the input matrix to the reservoir.

We have applied cross-validation and trained the reservoir five times with different combinations of the 5 dataset partitions, and tested the reservoir using the entire dataset to calculate the performance of both training and testing separately. In addition, statistics were collected by repeating this 10 times with different random seeds for constructing the reservoir. From these experiments, the percentage of correct utterance for training and testing were obtained and plotted as box plots in Figure~\ref{fig:enter-labelRD1} and \ref{fig:enter-labelRDf} .

\subsection{ RC-based MFCC Feature Extraction in the Time Domain}
\label{section:experiment_3}
 
\begin{figure}
   % \centering
    \includegraphics[width=.42\textwidth]{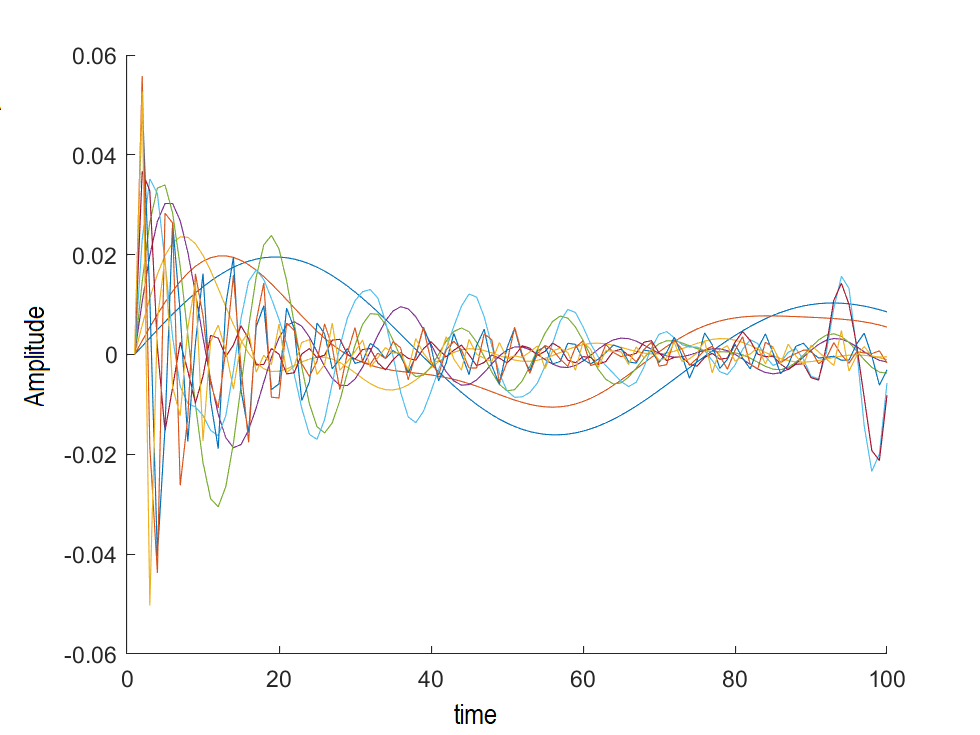}
    \caption{Time-domain filterbank}
    \label{fig:_FB}
\end{figure}

 Building upon the demonstrated classification capabilities of our reservoir computing (RC) system, we now establish its capacity for time-domain audio pre-processing. We have created the time domain filter bank signal corresponding to each Mel coefficient. For each Mel coefficient there is a set of frequencies and a set of parameters as shown in Table~\ref{tab:my_label}, where the frequency and parameters for first Mel coefficient are given. We have synthesised a sine wave corresponding to each frequency and parameter, and superimposed all the synthesized sine waves to get a Mel filter bank signal. Similarly synthesized all the Mel filter bank signals. Figure \ref{fig:_FB} shows the Mel filterbank in time domain.

In order to obtain MFCC in the time domain, the audio signal is convoluted with each of the Mel filterbank signals to obtain the corresponding Mel coefficient. We train the reservoir to perform convolution of the audio signal and the time domain filter bank signal.

\begin{table}
    \centering
    \begin{tabular}{@{}cccccccccc@{}}
    \toprule
         parameter= & 0.002454697 & 0.004909393 & 0.00736409& 0.009818787 \\
         %\hline
         $\Delta f$= & 131 & 141 & 151 & 161 \\
         \midrule
         parameter= &0.007781114 & 0.00561907 & 0.003457026 & 0.001294982\\
         %\hline
         $\Delta f$= & 171 & 181 & 191 & 201 \\
         %\hline
         \bottomrule
    \end{tabular}
    \vspace{3px}
    \caption{Parameters and frequencies for 10 Mel frequencies used}
    \label{tab:my_label}
\end{table}

\begin{figure}[h]
    \includegraphics[width=.52\textwidth]{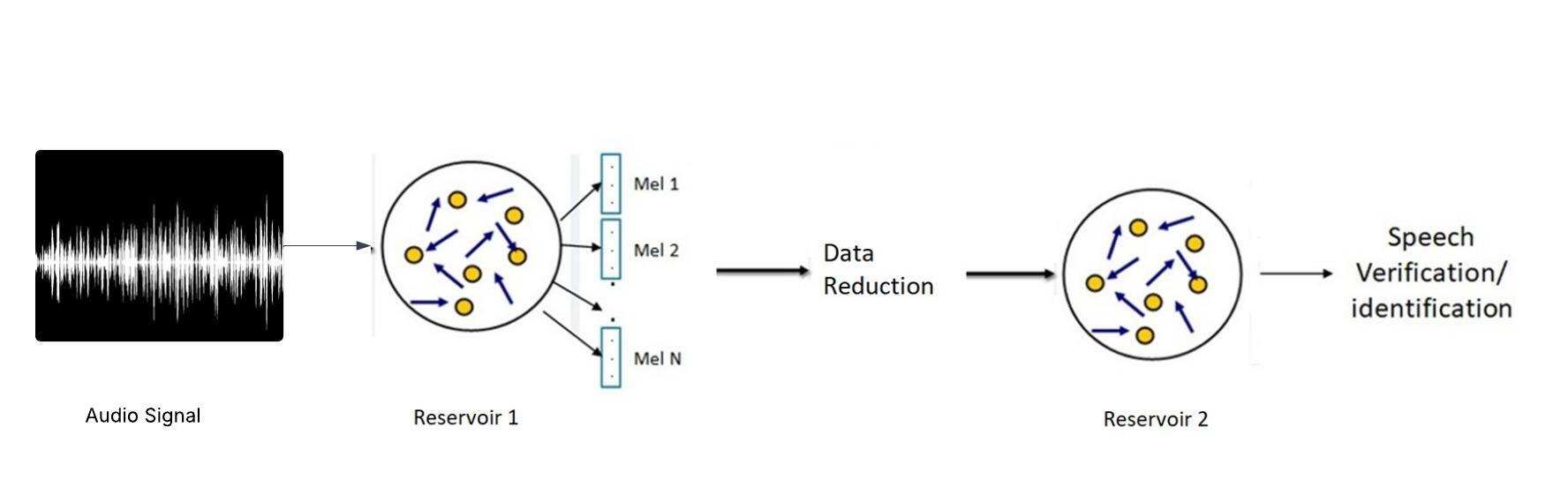}
    \caption{Time-domain MFCC extraction}
    \label{TM}
\end{figure}

Now the challenge is to reduce the number of data points without losing too much information. From the convoluted signal, we first trim the data-points which are beyond the length of the audio signal, because this part carries less information about the audio signal and is a residue of convolution operation. Now we split the signal into windows where the number of windows is equivalent to the number of MFCC output samples that we get when using the Matlab MFCC function in order to fit into the experiment framework with the second reservoir later. Out of each window we pick one data-point using absolute max pooling technique, i.e., the largest value. Figure \ref{TM} shows the MFCC extraction in time domain.
%\subsubsection{Setup 1: 1 reservoir 14 output layers}

 We have used a reservoir to obtain the time domain MFCC in two different ways: in the first approach, we have used a reservoir for convoluting the audio signal with each component of the Mel filter bank, and concatenated the output of convolution. This signal is then windowed and max-pooled to obtain the time domain MFCC. In the second approach, we first windowed the signal in such a way that each window provides one Mel coefficient. In each window we make use of the reservoir to convolute the windowed audio signal with the Mel filterbank, followed by slicing and max-pooling to obtain one Mel time domain coefficient.

 In comparison to approach two, we discovered that approach one is quicker and produces a smaller normalized mean square error (NRMSE). Therefore, we decide to use approach one. (Our results and discussions are based on approach one.)
%\subsubsection{Setup 2: 14 reservoirs }
 %We have used 14 reservoirs to obtain the each coefficient of time domain MFCC.

 \section{Results and Discussion}
 
To evaluate the performance of the proposed methods we have formulated two experiments. In Experiment 1 we are training the classifier reservoir with the MFCC obtained using Matlab function. To evaluate the reservoirs capability for extracting MFCC in time domain, we have formulated Experiment 2

\begin{table}[htbp]
    \centering
    \begin{tabular}{@{}l l l@{}}
    \toprule
         Experiment &Reservoir & Training and Testing Function \\
         \midrule
        Experiment 1 (Exp1) & 2&Matlab MFCC\\
       % Experiment 2 (Exp2) & 1, 2& Reservoir mimicking Matlab MFCC\\
        Experiment 2 (Exp2) &1, 2& Time domain MFCC\\
        
         \bottomrule
    \end{tabular}
    \vspace{3px}
    \caption{Experiments}
    \label{tab:my_labelt2}
\end{table}

\begin{table}[htbp]
    \centering
    \begin{tabular}{|l|l|}
    \hline
    \textbf{Models} & \textbf{Accuracy (\%)} \\ \hline
    LSM \cite{Verstraeten2005-yz} & 94.0 \\ \hline
    Liquid-SNN \cite{Srinivasan2018-tb} &77.7 \\ \hline
    Reservoir Computing (MEMS)\cite{Dion2018-rr}	&78\\ \hline
   Reservoir-based(Our method(EXp 1)) & 92.9  \\ \hline
    Reservoir-based(Our method(EXp 2)) & 91.82 \\ \hline\\
    \end{tabular}\\
    \caption{{Comparison of Performance of models with Ti-46 dataset for digit recognition}}
    \label{tab:Ti}
\end{table}
\begin{table}[h!]
    \centering
    \begin{tabular}{|l|l|}
    \hline
    \textbf{Models}  & \textbf{Accuracy (\%)} \\ \hline
    CNN \cite{Sridhar2023-yx} & 96.4 \\ \hline
    LSTM\cite{Sridhar2023-yx} & 95.23 \\ \hline
     AudioNet(Deep-NN) \cite{Becker2018-ni} & 92.53 \\ \hline
    Liquid-SNN \cite{Srinivasan2018-tb} &82.65 \\ \hline
    Reservoir-based(Our method(EXp 1)) & 93.08  \\ \hline
    Reservoir-based(Our method(EXp 2)) & 84.81 \\ \hline\\
    \end{tabular}
    \caption{Comparison of Performance of models with Audio-Mnist dataset for digit recognition.}
    \label{tab:Mnist}
\end{table}

\begin{figure*}
    \centering
   \subfloat[Speaker Recognition Performance using Audio-Mnist dataset]
    {\includegraphics[width=\columnwidth]{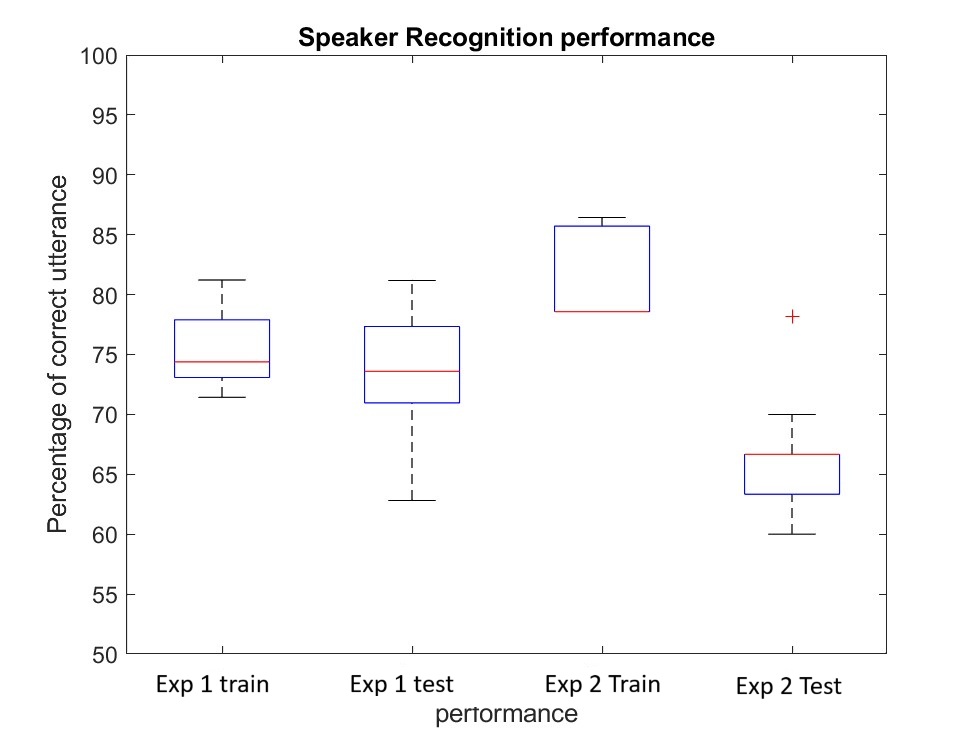}}
    \subfloat[Speaker Recognition performance using Ti-46 dataset]
     {\includegraphics[width=\columnwidth]{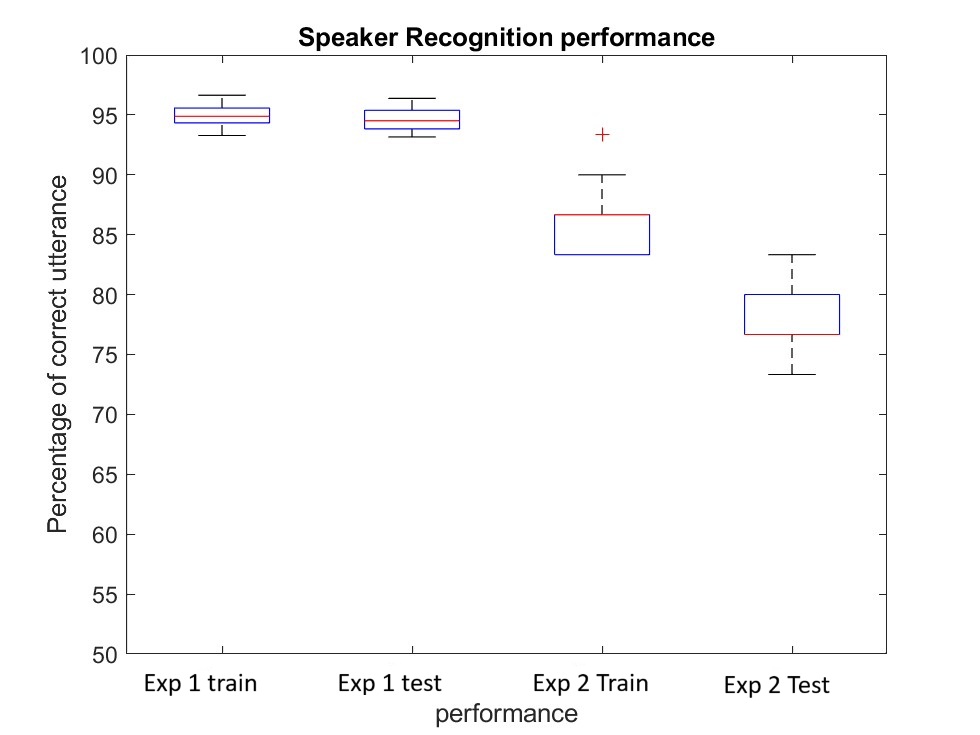}}
            \caption{Speaker Recognition performance of our system }
    \label{fig:enter-labelRD1}
\end{figure*}

\begin{figure*}
    \centering
   \subfloat[Digit Recognition performance using Audio-Mnist dataset]
    {\includegraphics[width=\columnwidth]{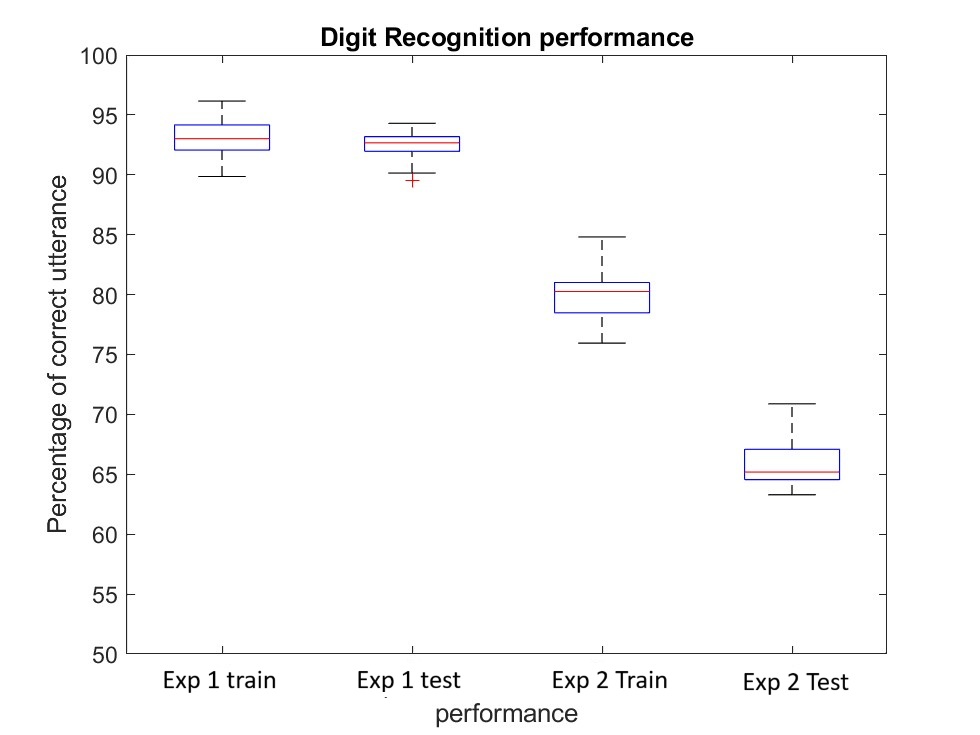}}
    \subfloat[Digit Recognition performance using Ti-46 dataset]
     {\includegraphics[width=\columnwidth]{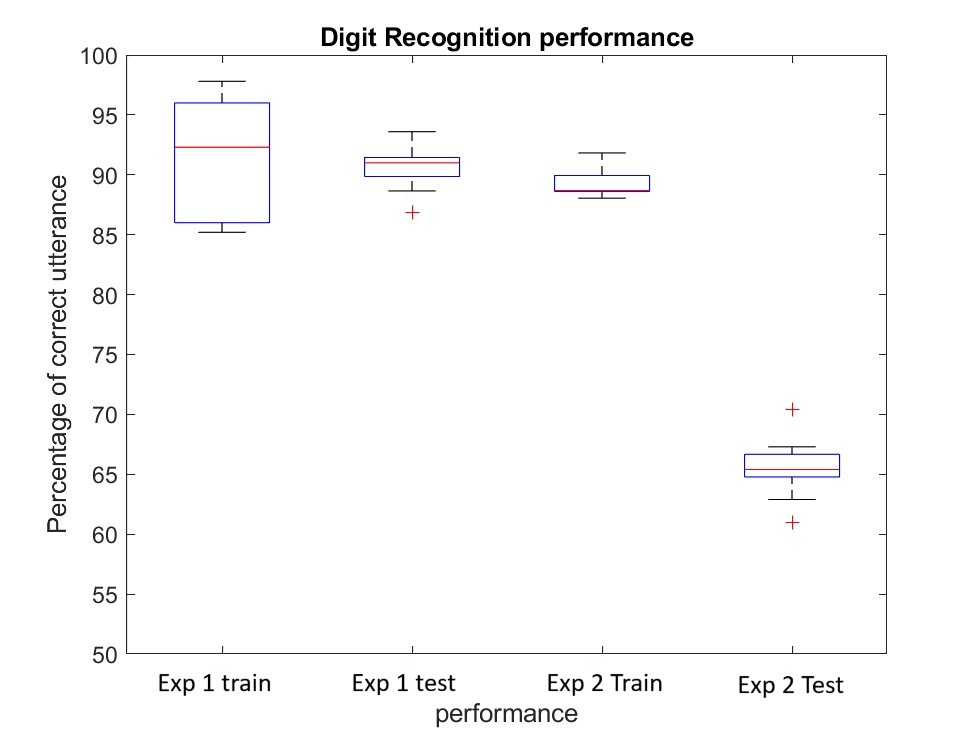}}
            \caption{Digit Recognition performance of our system}
    \label{fig:enter-labelRDf}
\end{figure*}

The box plots of training and testing performance of the reservoir as a classifier are shown in Figure~\ref{fig:enter-labelRD1} and \ref{fig:enter-labelRDf}.  Results for the case where MFCC are obtained from Matlab function are labeled \textit{Exp 1}, and the training and test performance of the reservoir which uses MFCC values obtained from the first reservoir trained to produce time domain MFCC  is labeled \textit{Exp 2}. The tables \ref{tab:Ti} and \ref{tab:Mnist} shows a comparison of the performance of different audio signal processing methods using Ti-46 and Audio-Mnist datasets respectively for digit Recognition.

From figure~\ref{fig:enter-labelRD1} and \ref{fig:enter-labelRDf} labeled Exp 1 we can see how well the reservoir, which uses the MFCC Matlab function performs. Output of Exp 1 demonstrates the reservoir's capacity to carry out classification and regression tasks. Given the complexity of audio analysis, any neural network that does audio classification often needs a large number of nodes, which takes time and energy to complete. In this instance, the classifier reservoir uses the least amount of energy, resources, and time while having just 100-400 nodes.
\\
\begin{table*}[htbp]
    \centering
    \begin{tabular}{@{}lcccccccccc@{}}
    \toprule
      % NRMSE of reservoir-1 mimicing MFCC(10 mel coefficents) \\
        %\hline
         Experiment & Mel 1 & Mel 2 & Mel 3& Mel 4& Mel 5& Mel 6& Mel 7& Mel 8& Mel 9& Mel 10 \\
         \midrule
         % Reservoir trained on Matlab MFCC &           0.8379&0.8423&0.8330&0.8937&0.9253&0.8824&0.9178&0.9409&0.9390&0.9597\\
%          Reservoir trained on Matlab MFCC &           0.83792&0.84229&0.833&0.89371&0.92528&0.88244&0.91775&0.94088&0.93902&0.95965\\
          Reservoir trained on Time-domain MFCC & 0.9757&0.9995&0.9527&0.9527&0.9557&0.9528&0.9093&0.7671&0.5075&0.6470\\ %&0.6587&\\
         \bottomrule
    \end{tabular}
    \vspace{3px}
    \caption{NRMSE of Reservoir 1 performance mimicking MFCC (10 Mel coefficients)}
    \label{tab:my_label1}
\end{table*}

\begin{table*}[h!]
    \centering
      \begin{tabular}{|c|c|c|c|}
        \hline
        \textbf{Models}  & \textbf{Train accuracy\%.} & \textbf{Test accuracy\%}  & \textbf{ Average number of neurons} \\
        \hline
        CNN  & 100 & 98.63 &~2M-10M parameters \\ \hline
       % Naive Bayes  & 94 & 94 ~100K parameters \\ \hline
        Word embedding  & 95.50 & 92.20 &~1M-5M parameters \\ \hline
        Logistic regression  & 64.32 & 61.95 &- (Depends on dataset size) \\ \hline
        Naive Bayes  & 50.25 & 49.75 &~100K parameters  \\ \hline
        SVM  & 82.88 & 83.32 &(Depends on support vectors) \\ \hline
        Random forest classifier VGG16 & 72.42 & 71.90 &~10K-100K trees \\ \hline
        ResNet50  & 91.30 & 80.20 &~5M parameters \\ \hline
        CapsNet  & 91.80 & 88.76 &~10M-20M parameters \\ \hline
        2D ConvNet bidirectional GRU & 68.85 & 65.23 &~10M-20M parameters \\ \hline
        Acoustic model  & 75.69 & 73.23 &(Depends on dataset size)\\ \hline
        CNN LSTM  & 83.25 & 80.52 &~5M-15M parameters \\ \hline
        Logistic regression 1-vector [26] & 84.30 & 80.23 &1M parameters \\ \hline
        LSTM-CNN [29] & 70.21 & 68.33 &~5M-15M parameters\\ \hline
        RC based(Using Matlab MFCC(Exp 1)) & 98.05 & 96.3 &400 Neurons \\ \hline
         RC based(Reservoir Mimicking MFCC)(\cite{Sebastian2026-eo} \ & 94.87 & 85.89 &RC-1=950 neurons, RC-2=400 neurons  \\ \hline
          RC based(time domain MFCC(Exp 2) & 93.33 & 83.33 &RC-1=35 neurons, RC-2=400 neurons  \\ \hline\\
         \end{tabular}
         \\
     \caption{Performance measures obtained from various language identification techniques\cite{Sebastian2026-eo}\cite{Singh2021-kp}\cite{Anushka-ac}\cite{Deng2023-xq}}
    \label{tab:speech_recognition_methods}
\end{table*}
%In the second experiment we are utilizing a reservoir to mimic the MFCC extraction. The normalized mean square error of reservoir 1, which is trained to mimic the Matlab MFCC, is displayed in Table~\ref{tab:my_label1}. This demonstrates the reservoir's capability to perform this task. The NRMSE value clearly emphasizes the ability of a reservoir to mimic MFCC extraction. We utilize the output of this first reservoir as the input to the second reservoir, and the resulting output is shown as box plots in figure~\ref{fig:enter-labelRD1} and \ref{fig:enter-labelRDf} labeled exp2. The plot shows  the reservoir's capability in mimicking MFCC feature extraction. In terms of speaker recognition, the outcomes of our method is are comparable to the results of Matlab based MFCC. 

Table~\ref{tab:my_label1} shows how well a reservoir is able to extract MFCC in the time domain. For the case where the time domain MFCC is applied to the input of the second classifier reservoir, the resulting performance is shown as box plots labeled Exp 2 in Figure~\ref{fig:enter-labelRD1} and \ref{fig:enter-labelRDf}. The plot shows  the ability of reservoir in extracting MFCC feature in time domain. Even with the simple max-pooling/averaging method that we have used to reduce the number of data points we are able to get good performance. We were able to get better results with methods like wavelet transform if used as the data reduction method; however, they complicate the system and detract from the main objective of our study. Our focus for future work is therefore to further develop a simple methodology to reduce the number of data points without losing relevant information of the signal.

Table~\ref{tab:speech_recognition_methods} shows a comparison of the number of neurons used by different audio signal processing methods. The number of neurons in a network is calculated based on architecture implementation and hyper parameters such as the number of hidden layers, the number of units (neurons) in each layer, and the input/output dimensions.  As can be seen from the table, Our method is the most lightweight and effective model, achieving high accuracy with smaller number of  neurons. By utilizing significantly less parameters while maintaining the MFCC extraction in the time domain, our method demonstrates good performance.

\section{Future development}

We need to further investigate and find a best fit reservoir 1 to improve overall performance of the spoken digit recognition system. Our future study will also focus on developing a simple method that reduces the number of data points without compromising signal information for time domain MFCC extraction.

In experiment 2 we are able to improve the performance of speaker recognition compared to experiment 1. Digit recognition is more challenging than speaker recognition. In short, our approach to MFCC extraction is simple, effective and promising. Our next objective will be to optimize the reservoir configuration. Additionally, we want to create a different technique that uses a reservoir to extract MFCC or any other superior speech feature.

\section{Conclusion}
    \begin{figure}
    \centering
    \includegraphics[width=.5\textwidth]{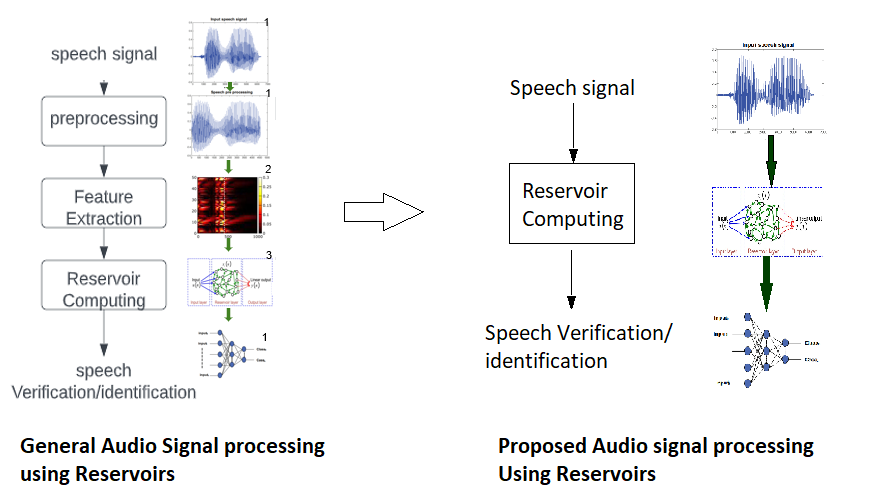}
   \caption{Audio processing using Reservoir computing}
    \label{fig:my_labelAP}
\end{figure}

Our aim is to develop an efficient audio processing system that works directly on audio samples in the time domain. Our method has extracted time-domain MFCC feature using a lightweight reservoir.

Figure~\ref{fig:my_labelAP} shows general audio signal processing alongside our proposed end-to-end reservoir system, where audio signals are fed directly into a first reservoir, another (or ultimately the same) reservoir processes the signal, and finally we obtain the desired output from a reservoir. The results shown in this paper demonstrate that a baseline end-to-end reservoir processing system has been successfully applied to audio signal processing in the time domain.

Our experimental findings shows that these method are feasible, as the reservoir-based system achieves competitive performance with a significant reduction in computational overhead. This achievement highlights the possibility of using time-domain reservoir computing as a low-complexity substitute for traditional frequency-domain techniques. Future research will focus on enhancing accuracy and optimizing reservoir topologies for wider audio processing applications.
\section{Data Availability Statement}
The original contributions presented in this study are included in the article. Further inquiries can be directed to the corresponding authors.
\section{Conflict of Interest}
The authors declare no conflict of intrest.
\bibliographystyle{plain}
\bibliography{Reference}

\end{document}